\documentclass[useAMS,usenatbib]{mn2e}
 \usepackage{graphicx}
\usepackage{natbib}
\usepackage{times}
\usepackage{amsmath}
\usepackage{amssymb}
\usepackage{textcomp}
\usepackage[normalem]{ulem}
\usepackage{multirow}
\usepackage{capt-of}

\title[Galaxy sizes  as a function of environment at intermediate $z$]{Galaxy sizes  as a function of environment at intermediate redshift from the ESO Distant Cluster Survey}
\author[Kshitija Kelkar et al.]{Kshitija Kelkar$^{1}$\thanks{E-mail:
ppxkk1@nottingham.ac.uk (KK) }, Alfonso Arag\'{o}n-Salamanca$^{1}$, Meghan E. Gray$^{1}$, David Maltby$^{1}$, \newauthor Benedetta Vulcani$^{2}$, Gabriella De Lucia$^{3}$,
Bianca M. Poggianti$^{4}$, Dennis Zaritsky$^{5}$  \\
$^{1}$School of Physics \& Astronomy, University of Nottingham, Nottingham NG7 2RD, UK\\
$^{2}$Kavli Institute for the Physics and Mathematics of the Universe (WPI), Todai Institutes for Advanced Study, University of Tokyo, Kashiwa 277-8582, Japan\\
$^{3}$INAF-Osservatorio Astronomico di Trieste, Via Tiepolo 11,I-34131 Trieste, Italy\\
$^{4}$INAF-Osservatorio Astronomico di Padova, Vicolo dell Osservatorio, 5, I-35122, Padova, Italy\\
$^{5}$University of Arizona, 933 N. Cherry Ave, Tucson, AZ 85721, USA\\
  }

\begin{document}
\date{Accepted xxx. Received xxx; in original form xxx}

\pagerange{\pageref{firstpage}--\pageref{lastpage}} \pubyear{xxxx}

\maketitle

\label{firstpage}

\begin{abstract}

{In order to assess whether the environment has a significant effect on galaxy sizes, we compare the mass--size relations of cluster and field galaxies in the $0.4 < z < 0.8$ redshift range from the ESO Distant Cluster Survey (EDisCS) using HST images. We analyse two mass-selected samples, one defined using photometric redshifts ($10.2 \le \log M_\ast/M_{\odot} \le 12.0$), and a smaller more robust subsample using spectroscopic redshifts ($10.6 \le \log M_\ast/M_{\odot} \le 11.8$). We find no significant difference in the size distributions of cluster and field galaxies of a given morphology. Similarly, we find no significant difference in the size distributions of cluster and field galaxies of similar rest-frame $B-V$ colours. We rule out average size differences larger than $10$--$20$\% in both cases. Consistent conclusions are found with the spectroscopic and photometric samples. These results have important consequences for the physical process(es) responsible for the size evolution of galaxies, and in particular the effect of the environment. The remarkable growth in galaxy size observed from $z\sim2.5$ has been reported to depend on the environment at higher redshifts ($z>1$), with early-type/passive galaxies in higher density environments growing earlier. Such dependence disappears at lower redshifts. Therefore, if the reported difference at higher-$z$ is real, the growth of field galaxies has caught up with that of cluster galaxies by $z\sim1$. Any putative mechanism responsible for galaxy growth has to account for the existence of environmental differences at high redshift and their absence (or weakening) at lower redshifts. }
    
\end{abstract}


\section{Introduction}
\label{secintro}

It is well established that galaxies have shown remarkable evolution in their physical properties such as their sizes over cosmic time. Observations of present-day galaxies clearly show that their sizes are correlated with their stellar masses, and
that this correlation evolves significantly over look-back time. This size evolution was put forth by some of the first works published on this topic showing that massive quiescent 
galaxies at high redshift ($z > 1$) were much more compact than their local counterparts \citep{vdokkum08,daddi05, buitrago08, shen03, mc05}. This growth in galaxy sizes was found to be particularly dependent on morphology at a given stellar mass \citep{trujillo06,trujillo07}: spheroids were four times more compact at $z\sim1.5$ than their local counterparts, whereas $z\sim1.5$ disk-like galaxies were twice as compact as $z\sim0$ ones. Also it was observed that such 
size evolution was mostly observed in massive galaxies with low star-formation rate \citep{franx08}.

Several processes are being considered as possible drivers of the observed size evolution. Rapid mass loss of cold gas from the central regions due to AGN feedback \citep{fan08} could explain the size growth of massive spheroidal galaxies. Size evolution could also be triggered by environment-dependent processes such as `dry' mergers, resulting in the increase in size of the stellar distribution in early-type galaxies but without any further star formation due to lack of gas \citep{vdokkum05}. The accretion of stars following a 
merger could also lead to the growth of a galaxy's size due to an envelope being formed around the merger remnant \citep{naab07}.  

Hierarchical models of galaxy formation and evolution predict that the densest regions collapsed first and therefore cluster galaxies would form earlier compared with field galaxies of similar mass \citep{dlucia04}. In essence, cluster galaxies would have had a `head-start' in their evolution as compared to field galaxies. These predictions seem to have some observational support. \cite{lani13} find that the mass--size relation of quiescent galaxies at $1\leq z\leq 2$ shows some environmental dependence: galaxies in dense environments appear to be larger than those in low-density regions, suggesting that the galaxy growth may have happened earlier in the densest environments. This dependence gets weaker from $z\sim 1$ to $z\sim 0.5$, indicating, perhaps, that size evolution in the low-density regions is `catching up' with that in denser ones. 
However, it is not clear whether the observed dependency of size evolution with environment is due to the fact that galaxies in high-density environments are, on average, older or if it is due to environment-driven processes. 

Furthermore, \citet{hcompany13} find no environmental dependence of the mass--size relation between $z\sim 1$ and the present. There is also extensive evidence indicating that at $z\sim0$ the mass--size relation for galaxies with a given morphology does not depend significantly on environment either \citep{shen03, maltby10, rettura10, poggianti13}, although some subtle differences may still be present \citep{cebrian14}. This seems to suggest that there must be an epoch when the environment ceases to affect galaxy sizes, perhaps because the growth has already been completed in all environments.

There are, however, very significant difficulties when comparing different studies at different redshifts. The details governing the sample selection (e.g., galaxy mass and redshift range), sample division (e.g., by galaxy colour, visual morphology, S\'ersic index, quiescence) and the exact definition and measurement of the environment are critical. Subtle (and not-so-subtle) differences make direct comparisons dangerous. Some of these difficulties may be behind the discrepancies found in different studies. For instance, \citet{raichoor12} used morphologically-selected early-type galaxies at $z\sim1.2$ and found that the high mass ($10 < \log (M_\ast/M_{\odot}) < 11$) cluster galaxies are significantly more compact than similar field galaxies, whereas \cite{cooper11} and \cite{papovich12} found an opposite trend, albeit at higher redshift. It is also important to remember that, as \citet{valentinuzzi10} found, progenitor bias plays a crucial role when introducing spurious size-evolution while comparing high- with low-redshift morphological samples.

It is clear that the picture still remains incomplete as to whether and, if so, when galaxy size evolution depends on environment. In this paper, therefore, we aim to examine the role of environment on the galaxy stellar mass--size relation in the $0.4<z<0.8$ redshift range since this could be the transition epoch when the putative environmental differences found at higher redshifts cease to be present. We will compare samples of cluster and field galaxies in a relatively narrow redshift range to avoid evolutionary effects so that we can concentrate on purely environmental ones. In order to do that we will construct mass-selected samples of galaxies, subdivided both by colour and morphology, in cluster and field environments taken from the ESO Distant Cluster Survey \citep[EDisCS,][]{white05}.

This paper is structured as follows: Section~\ref{secdata} describes the data, the sample selection, and the methodology used when defining the environment and computing the galaxy sizes. Section~\ref{secanalysis} presents the analysis and discussion of the mass--size relations for each of our samples. Finally, in Section~\ref{secconclusions} we present our conclusions.
Throughout this paper, we use the standard $\Lambda$CDM Cosmology with $h_{0} = 0.7$, $\Omega_{\Lambda} = 0.7$ and  $\Omega_{\rm m} = 0.3$.

\section[]{Description of the data}
\label{secdata}

The ESO Distant Cluster Survey (EDisCS) is a multiwavelength survey comprising 20 fields containing galaxy clusters in the redshift range $0.4<z<1$ \citep{white05}. The fields were originally drawn from the parent catalogue derived from the Las Campanas Distant Cluster Survey \citep[LCDS; ][]{gonzales01}. Optical photometry for all fields was obtained using FORS2 on the ESO Very Large Telescope (VLT),
details of which are given in \citet{white05}. This ground-based photometry included $B$-, $V$-, and $I$-band imaging for the ten intermediate-$z$ ($0.4<z<0.5$) clusters and $V$-, $R$- and $I$-band imaging for the remaining ten high-$z$ ($0.5<z<1$) clusters. Spectroscopy with FORS2/VLT was carried out on galaxies selected using the $I$-band magnitude and the best-fit photometric redshift from the photometric sample, as described in \citet{halliday04} and \citet{mjensen08}. {The selection criteria applied ensure that the spectroscopic sample is, essentially, an $I$-band selected sample.} Please refer to the above papers for a general description of the EDisCS clusters. 

In addition, the ten high-$z$ clusters also have HST $I$-band imaging data taken using the $F814$ filter on the ACS Wide Field Camera in a total of five pointings per field: four adjacent 1-orbit pointings covering $6.5^\prime \times 6.5^\prime$ (which approximately matches the field of the ground-based VLT optical images) and an additional 4-orbit pointing covering the central $3.5^\prime \times 3.5^\prime$ region of each cluster, centred at the location of the BCG \citep{desai07}.
The resulting exposure time for the cluster centres was therefore $10200\,$seconds, whereas the surrounding area had an effective exposure of $2040\,$seconds. {These 10 cluster fields with HST data will be the ones analysed in this paper since HST images are needed to obtain accurate galaxy morphologies and sizes.    
Table~\ref{table1} gives a summary of the properties of the cluster sample used in our analysis.} 

Note that since the field galaxies are selected and observed in exactly the same way as those in the clusters, they provide a field galaxy sample that can be directly and reliably compared with the cluster sample.  

\begin{table}
\caption{Summary of the cluster sample properties (including secondary clusters, cf. \S\ref{subsecenvironment}). Columns~1--5 contain the cluster ID,  cluster redshift, cluster velocity dispersion, cluster mass, and the number of spectroscopically confirmed cluster members \citep{halliday04,mjensen08}. The cluster masses have been estimated from the velocity dispersions following \citet{finn05}.  \label{table1}}
\begin{center} 
\begin{tabular}{|l|c|c|c|c|}
 
 \hline
 Cluster &  $z_{\rm cl}$ & $\sigma_{\rm cl}$ & $\log M_{\rm cl}$ & No.\ of spec. \\
 &  & (km$\,$s$^{-1}$) &  ($M_\odot$) & members\\
 \hline
 cl1037$-$1243 & 0.5783 & 319$^{+53}_{-52}$ & 13.61 & 16 \\
 cl1037$-$1243a & 0.4252 & 537$^{+46}_{-48}$ & 14.33 & 43 \\
 cl1040$-$1155  & 0.7043 & 418$^{+55}_{-46}$ & 13.93 & 30 \\
 cl1054$-$1146  & 0.6972 & 589$^{+78}_{-70}$ & 14.38 & 48 \\
 cl1054$-$1245  & 0.7498 & 504$^{+113}_{-65}$ & 14.16 & 35 \\
 cl1103$-$1245 & 0.9586 & 534$^{+101}_{-120}$ & 14.18 & 9 \\
 cl1103$-$1245a  & 0.6261 & 336$^{+36}_{-40}$ & 13.66 & 14 \\
 cl1103$-$1245b  & 0.7031 & 252$^{+65}_{-85}$ & 13.27 & 11 \\
 cl1138$-$1133  & 0.4796 & 732$^{+72}_{-76}$ & 14.72 & 45 \\
 cl1138$-$1133a  & 0.4548 & 542$^{+63}_{-71}$ & 14.33 & 11 \\
 cl1216$-$1201  & 0.7943 & 1018$^{+73}_{-77}$ & 15.06 & 66 \\
 cl1227$-$1138  & 0.6357 & 574$^{+72}_{-75}$ & 14.36 & 22 \\
 cl1227$-$1138a  & 0.5826 & 341$^{+42}_{-46}$ & 13.69 & 11 \\
 cl1232$-$1250  & 0.5414 & 1080$^{+119}_{-89}$ & 15.21 & 52 \\
 cl1354$-$1230  & 0.7620 & 648$^{+105}_{-110}$ & 14.48 & 20 \\
 cl1354$-$1230a  & 0.5952 & 433$^{+95}_{-104}$ & 14.00 & 14 \\
 \hline
\end{tabular}
\end{center}
\end{table}

\subsection[]{Sample selection}
\label{subsecsample}

In order to explore the effect of the environment on galaxy sizes we need to build stellar-mass-selected samples of cluster and field galaxies. These samples will then be split up by galaxy morphology and colour and their mass--size relations compared.

Galaxies are allocated to individual clusters and the general field using both spectroscopic and photometric redshifts. Galaxies with spectroscopic redshifts form a subset of the EDisCS photometric sample. {These spectroscopic target galaxies were selected using their $I$-band magnitudes measured in a $1^{\prime\prime}$-radius circular aperture, with the mid--$z$ and high--$z$ fields having apparent magnitude limits of 18.6 and 19.5 respectively \citep{mjensen08}. The spectroscopic sample contains relatively bright galaxies, and therefore surface-brightness selection effects are negligible because the FORS2 images reach much deeper than the spectroscopic limits (by 3–-4 magnitudes). As discussed in \citet{mjensen08}, the spectroscopic sample is effectively an $I$-band limited sample.  }   
We will explore the spectroscopic subsample first, which is obviously smaller but more robust, with more reliable distances (and thus sizes and stellar masses) and cluster membership information. Later, we will explore the photometric sample, much larger but with reduced reliability.

The stellar masses for the spectroscopic sample and photo-$z$ cluster members (see below) were computed by Benedetta Vulcani (2013, private communication). Distances to the galaxies were calculated using the individual spectroscopic redshifts for field galaxies and the mean cluster redshift \citep{halliday04, mjensen08} for cluster members. The stellar masses for the spectroscopic sample and  photo-$z$ cluster members were computed using the \citet*{kroupa01} IMF following the method proposed by \citet*{bell01}. The mass completeness limit for the spectroscopic sample is $\log M_\ast/M_\odot = 10.6$ \citep{vulcani10} and for the photometric sample $\log M_\ast/M_\odot = 10.2$ \citep{vulcani11}. These mass completeness 
limits were obtained from the most distant cluster in our sample, cl1216.8$-$1201, by determining the mass of a galaxy with an absolute $B$-magnitude corresponding to $I=24(23)$ for the photometric(spectroscopic) sample and a
colour $B-V \sim 0.9$, which is the reddest colour of the galaxies in this cluster.   
The stellar masses for the photometric 
field galaxies were not computed as the photometric redshifts are not accurate enough to estimate reliable distances and thus the rest-frame luminosities and colours required to calculate stellar masses. In all our samples, we removed the BCGs from each of the ten clusters, as identified by \citet{whiley08}, since they do not follow the mass--size relation of normal galaxies \citep{maltby10}. Table~\ref{table} provides information on the spectroscopic and photometric samples used in this paper. Details on how these samples are defined are given below. Note that further refinement of these samples will be described in section~\ref{sec:sizedetermination}.

\subsubsection[]{Morphologically selected sample}
\label{subsectmorphologies}

{Visual morphologies for the galaxies were obtained from the HST morphology catalogue published by \citet{desai07}. The morphological classifications were carried out by 5 classifiers, and the final morphology for each galaxy was assigned using a weighted combination of the individual classifications. This process is designed to minimize uncertainties and individual biases \citep{desai07}. For the spectroscopic morphology-selected sample, only galaxies with spectroscopic redshifts that were covered by the HST images were used. Furthermore, in this paper we have collapsed the fine morphological classes given by the original catalogue into three broad bins: Ellipticals, S0s and Spirals. Other morphological types (mostly irregulars) were excluded. Given the qualitative nature of the classification process, formal uncertainties are very difficult to assign to the morphologies,  but since the morphological classes used here are much coarser than those given in the original paper, the broad classes that we use are expected to be reasonably robust.} 
The spectroscopic morphology-selected sample contains 213 cluster galaxies and 167 field galaxies. The photometric morphology-selected sample was constructed using galaxies with known morphologies found in the photo-$z$ catalogues \citep{white05, pello09}, and contains 1167 cluster galaxies and 278 field 
galaxies\footnote{Please note that, as discussed in section~\ref{subsecenvironment}, when building the field sample for comparison with the photometric cluster sample, we used the spectroscopic field sample but with a stellar-mass limit of $\log M_\ast/M_\odot = 10.2$, the same limit used for the photometric cluster sample}. See Section~\ref{subsecenvironment} for details of how cluster membership was determined.

\subsubsection[]{Colour-selected sample}
\label{subsectcolour}

\begin{figure}
 \includegraphics[width=0.5\textwidth]{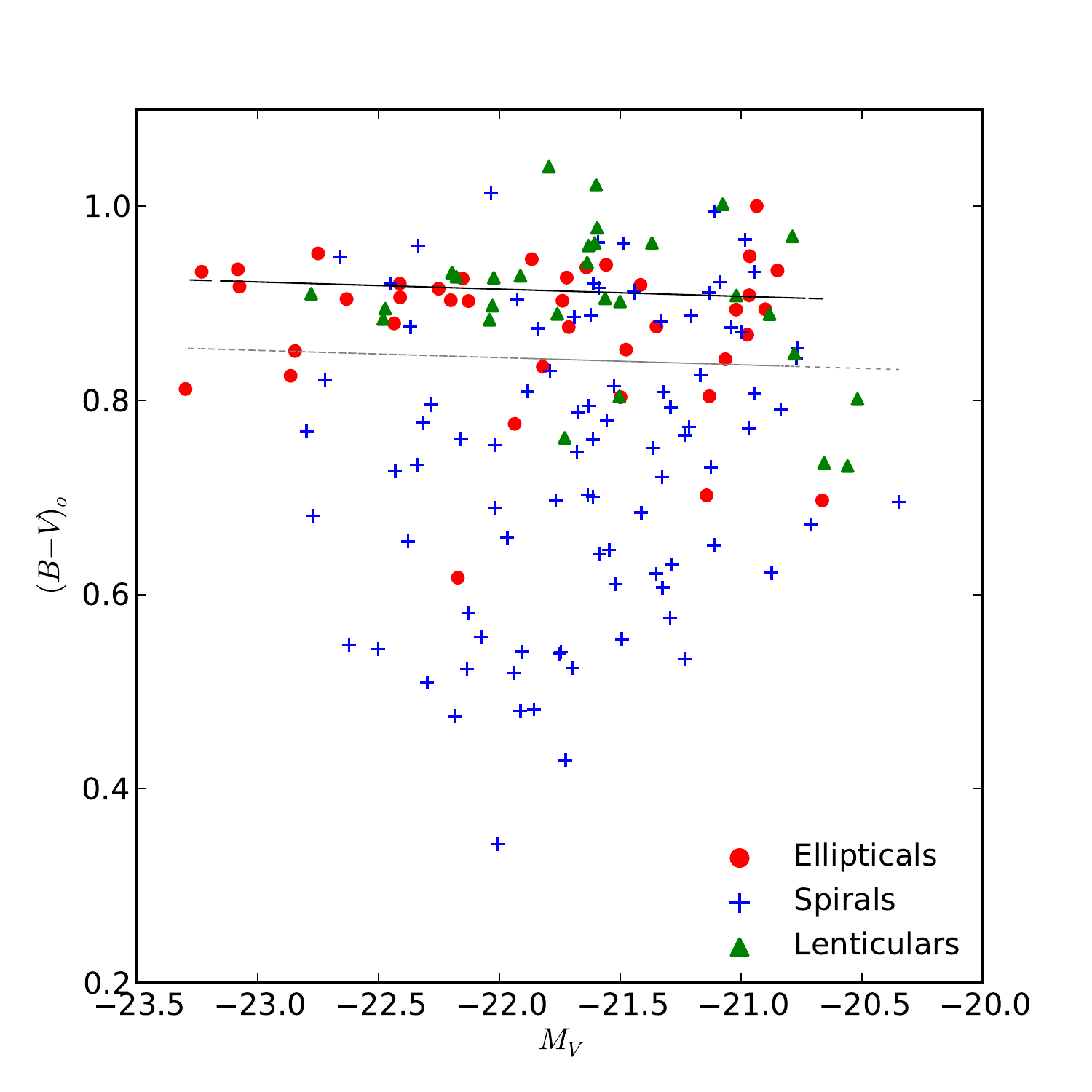}
 \caption{\label{figcmd} The rest-frame $B-V$ colour--magnitude relation for the combined spectroscopic cluster sample. The solid line shows a linear fit to the red sequence, while the dotted grey line indicates the boundary separating the red and blue galaxy samples. This boundary has the same slope as the red sequence but is located $3\sigma$ below, where $\sigma$ is the scatter of the elliptical galaxy colours around the red sequence.}
\end{figure}

Figure~\ref{figcmd} shows the rest-frame colour--magnitude diagram for the spectroscopic morphology cluster sample. These colours were computed using a $1^{\prime\prime}$ radius aperture for galaxies in crowded fields and the SExtractor ISO aperture \citep{bertin96} for other galaxies \citep{rudnick03}. Rest-frame colours were interpolated from the observed magnitudes using spectral energy distributions fitted using the observed spectroscopic redshifts \citep{pello09}. To split the sample by colour, a linear fit to the colour-magnitude relation of the elliptical galaxies was used to determine the red sequence location and slope. The boundary separating the red and blue galaxy samples was defined as a line with the same slope as the red sequence but located $3\sigma$ below, where $\sigma$ is the scatter of the elliptical galaxy colours around the red sequence. All the galaxies above this line are considered red and all below blue. The same boundary was used for the spectroscopic and photometric samples.

\begin{table*}
\caption{Properties and sizes of the different samples and sub-samples analysed in this paper. See text for details. \label{table}}
 \begin{tabular}{|l|c|c|c|c|}
 \centering
                &    \multicolumn{2}{c}{\bf Spectroscopic} &  \multicolumn{2}{c}{\bf Photometric}  \\
\hline
Mass range & \multicolumn{2}{c}{$10.6 \le \log \left(\frac{M_\ast}{M_{\odot}}\right) \le 11.8$}  &  \multicolumn{2}{c}{$10.2 \le \log \left(\frac{M_\ast}{M_{\odot}}\right) \le 12.0$}  \\
\hline
{\bf Parent Sample} &  & &  \\
\multicolumn{5}{c}{}\\
Cluster & \multicolumn{2}{c}{213}  & \multicolumn{2}{c}{1167} \\
Field   & \multicolumn{2}{c}{167}  & \multicolumn{2}{c}{278}  \\
\hline
{\bf Morphology}  & \multicolumn{2}{c}{\ E \ \ \ \ S0 \ \ \ Sp}  & \multicolumn{2}{c}{\ E \ \ \ \ \ S0 \ \ \ Sp}  \\
\multicolumn{5}{c}{}\\
Cluster     & \multicolumn{2}{c}{39 \ \ \ 29 \ \ \ 90} & \multicolumn{2}{c}{224 \ \ 116 \ \ 605}  \\
Field       & \multicolumn{2}{c}{21 \ \ \ 15 \ \ \ 83} & \multicolumn{2}{c}{\ \ 35 \ \ \ \ 22 \ \ 160}  \\
\hline
{\bf Colour} &  \multicolumn{2}{c}{Red \ \ \ \  Blue} & \multicolumn{2}{c}{Red \ \ \ \ Blue} \\
\multicolumn{5}{c}{}\\
Cluster     & \multicolumn{2}{c}{80 \ \ \ \ \ \ \ 78} & \multicolumn{2}{c}{ 272 \ \ \ \ \ \ 673}  \\
Field       & \multicolumn{2}{c}{46 \ \ \ \ \ \ \ 73} & \multicolumn{2}{c}{\ \ \ 54 \ \ \ \ \ 163}  \\
\hline
\end{tabular}
\end{table*}

\subsection[]{Environment definition}
\label{subsecenvironment}
In this paper, we define the global environment of the galaxies based on their cluster membership. The method used to determine cluster membership is different for the spectroscopic and photometric samples. Following \citet{mjensen08}, for the spectroscopic sample a galaxy is considered to belong to a cluster if its spectroscopic redshift lies within $\pm3\sigma$ from the average redshift of the cluster ($z_{\rm cl}$). {The spectroscopically-defined cluster membership is therefore expected to be very robust}.

For the photometric sample, cluster membership was determined following the method described in \citet{pello09}. Cluster members were selected by assigning them a probability of belonging to the cluster that had to be greater than a given threshold. This threshold was calibrated using the spectroscopic sample in order to minimize contamination and maximize completeness. Explicitly, the probability distribution function of each galaxy's photometric redshift is integrated within a $\pm0.1$ redshift slice about the cluster redshift $z_{\rm cl}$, and the value of this integral has to be higher than the empirically-determined threshold in order to consider a galaxy as cluster member. Extensive tests carried out by \citet{pello09} using the large number of available spectroscopic redshifts indicate that the method retains $\sim90$\% of the cluster members while rejecting $\sim88$\% of non-members for the clusters studied here. However, it is important to point out that these tests are based on the brighter spectroscopic sample. For the photometric sample these numbers are, strictly speaking, upper limits.  

All the galaxies not identified as cluster members are considered to be in the field sample. When building the field sample for comparison with the photometric cluster sample, we used the spectroscopic field sample but with a stellar-mass limit of $\log M_\ast/M_\odot = 10.2$, the same limit used for the photometric cluster sample. As mentioned in Section~\ref{subsecsample}, the photo-$z$ uncertainties prevent us from obtaining reliable distances (and thus stellar masses and intrinsic sizes) for field galaxies without spectroscopic redshifts. 

Note also that some of the EDisCS fields contain secondary clusters in addition to the main ones. These were discovered when analysing the spatial distributions and spectroscopic redshifts of the galaxies in each field \citep{white05, mjensen08}. Members of these secondary clusters are, for consistency, also included in the cluster sample. These secondary clusters are denoted in Table~\ref{table1} with `a' or `b' following the main cluster ID.

The field galaxies have a very similar redshift distribution to the cluster galaxies studied here. This ensures that a direct comparison can be made, avoiding redshift-dependent evolutionary effects.

\subsection[]{Galaxy size determination}
\label{sec:sizedetermination}
We used the data pipeline \textsc{GALAPAGOS} \citep[Galaxy Analysis over Large Areas: Parameter Assesement by \textsc{GALFIT}ting Objects from 
SExtractor;][]{barden12} to obtain the effective radii ($R_{\rm e}$) along the semi-major axis of the 2-D surface brightness distribution of each galaxy in the ten high-$z$ clusters using the HST I-band images. This pipeline uses \textsc{SE}xtractor \citep{bertin96} for source detection in the input images and \textsc{GALFIT} \citep{peng02,peng10} to fit a 2-D S\'ersic (1968) \nocite{sersic1968} $r^{1/n}$ model to the individual sources. This best-fit model is obtained by $\chi^{2}$-minimisation using a Levenberg-Marquardt algorithm\nocite{press1997}. \textsc{GALFIT} also determines several structural parameters along with the S\'ersic index $n$ and $R_{\rm e}$, but we will not discuss them here. We compared our results with those obtained using the \textsc{GIM2D} software \citep{simard02} and found that \textsc{GALAPAGOS} is much more robust in giving reliable fits in crowded regions as it simultaneously fits neighbours. A detailed analysis and comparison of both these codes is given in \citet{haussler07}.

To ensure that the sizes used in this study are robust, we visually inspected all the fitted models and the residuals to reject unreliable sizes. We found that fits yielding unphysical values of the parameters ($R_{\rm e} \geq 5^{\prime\prime}$ and/or $n < 0.2$ or $n > 6$) were almost always unreliable, leaving very significant residuals. The converse is also true: when the inspected residuals indicated that the fitted model was not a reasonable representation of the galaxies' surface brightness distribution, the fitted parameters were often unphysical. The reasons for this were varied. Some galaxies were affected by uncertain sky subtraction in very crowded regions. A few were very close to the edges of the images or had relatively bright stars nearby. There is also a significant number of galaxies showing strong perturbations caused by interactions or undergoing mergers. It is not surprising therefore that in all these cases a S\'ersic model fails to provide a good description of the light distribution of the galaxies. Since S\'ersic models are not suitable to fit these objects, the derived model parameters (including the effective radii) are meaningless and cannot be trusted. The fractions of field and cluster objects rejected this way were comparable ($\sim20$--$30$\%), and no obvious environment-depended biases were apparent. These objects were removed from the samples and not considered in our subsequent analysis. As a sanity check we tested that including them would not change our main conclusions.  It could be argued that since including these objects in our analysis does not change our conclusions, we could leave them in the sample. However, the advantage gained by increasing the sample size is lost because including these objects only adds noise to the statistical tests. On balance, we decided not to include them.  

Table~\ref{table} gives the numbers of galaxies in each subsample after eliminating unreliable fits. 
The measured $R_{\rm e}$ were converted into intrinsic linear sizes using the standard cosmology (cf. Section~\ref{secintro}), the individual spectroscopic redshifts for the field galaxies, and the relevant value of $z_{\rm cl}$ for the cluster members.

\section[]{Observed mass--size relations}
\label{secanalysis}

In this section, we compare the stellar mass--size relation for the field and cluster galaxies divided by morphology (Figures~\ref{figmsspecmorph} and~\ref{figmsphotmorph}) and rest-frame colour (Figures~\ref{figmsspeccol} and~\ref{figmsphotcol}) 
for the spectroscopic and photometric samples. In each figure we present the mass--size relation for the galaxies in each sample, together with averaged values in stellar-mass bins. These bins are arbitrary, and are only shown for illustration. The statistical analysis is carried out using the full cumulative distributions, also shown in the figures. 
Two-sample Kolmogorov--Smirnov (K--S) tests were carried out to estimate the probability $p$ that the field and cluster samples are derived from the same $R_{\rm e}$ distribution. Environmental differences on the mass--size relations were considered to be significant if $p<0.05$,
that is, when the significance is larger than $2\sigma$.   

Note that it makes no physical sense to compare the mass--size relations of all cluster and field galaxies disregarding their morphology or colour. The morphology/colour mix of the cluster population is very different from that of the field \citep[e.g.,][]{butcher78,dressler84,desai07}. Since early- and late-type galaxies have very different mass--size relations, any observed differences in the mass--size relations of the cluster and field global populations would be driven by differences in the morphology/colour mix, and therefore the samples need to be divided accordingly. Since different mass--size relation studies often use different criteria to divide the galaxy population into sub-populations, in order to assess whether these different criteria yield different results we decided to divide our sample both by morphology and colour. 

{It is very important to emphasise here that it is also necessary to separate elliptical and S0 galaxies in this analysis since they have very different profiles and formation histories. Indeed, \citet{bernardi13} and \citet{hcompany13} found significant differences in the mass--size relation for ellipticals and S0s. Our data confirm these findings: as figures~\ref{figmsspecmorph} and~\ref{figmsphotmorph} show, ellipticals are systematically larger (by $\sim20$--$25$\%) than S0s with similar masses.}

Before studying the effect of the environment on galaxy sizes, it is important to check that the stellar mass distribution of the galaxies in the subsamples we compare are similar. \cite{vulcani13} found that the mass distributions for red and blue galaxies do not significantly depend on the global (i.e., cluster vs.\ field) environment at redshifts comparable to ours. Furthermore, \cite{calvi13} found that at lower redshifts the variations of the mass functions with global environment for galaxies with similar morphologies are quite small and subtle, and possibly only affect galaxies with the highest masses\footnote{Note that we have eliminated the brightest cluster galaxies from our sample.}. We found very similar results. Two-sample K--S tests were used to look for differences in the mass distributions of galaxies in cluster and field environments, both separated by morphology and by colour. We found that the mass distributions for red and blue galaxies do not significantly depend on their global environment. Similarly, dividing the samples by morphology we found that the K--S tests failed to detect significant differences in the mass distributions of field and cluster galaxies. We are therefore confident that the mass distributions of the galaxy samples we compare are similar in clusters and in the field.

\subsection[]{Mass--size relation for morphologically-selected samples}
\label{subsecmsmorph}

\begin{figure*}
 \includegraphics[width=1.1\textwidth]{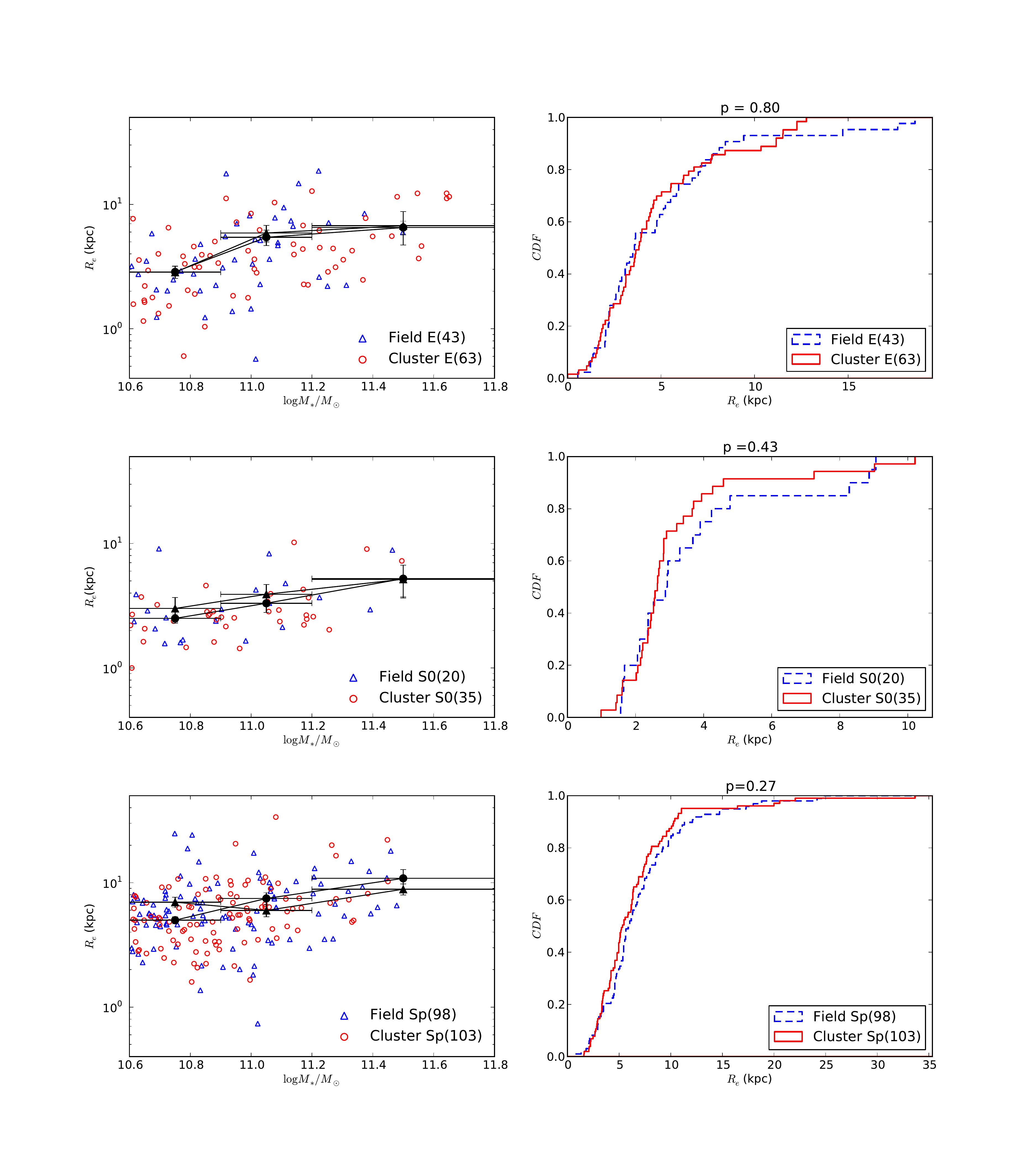}
 \caption{\label{figmsspecmorph} Stellar mass--size relation for the spectroscopic morphology-selected samples. {\textit{Left:}} The stellar mass--size relations for Spiral, Elliptical and S0 field (blue triangles) and cluster (red circles) galaxies. The larger black triangles and circles show average values in arbitrary mass bins for the field and cluster galaxies respectively. The horizontal error bars span the size of each bin, and the vertical error bars correspond to $1\sigma$-errors in the mean $R_{\rm e}$ of each bin.
{\textit{Right:}} The corresponding $R_{\rm e}$ cumulative distribution function (CDF) for each of the morphologies. The dashed blue and red lines correspond to the field and cluster samples respectively.   The numbers above each CDF plot correspond to the $p$ value derived form the K-S tests discussed in the text. The numbers in the boxes denote sample sizes. }
\end{figure*}

\begin{figure*}
 
 \includegraphics[width=1.1\textwidth]{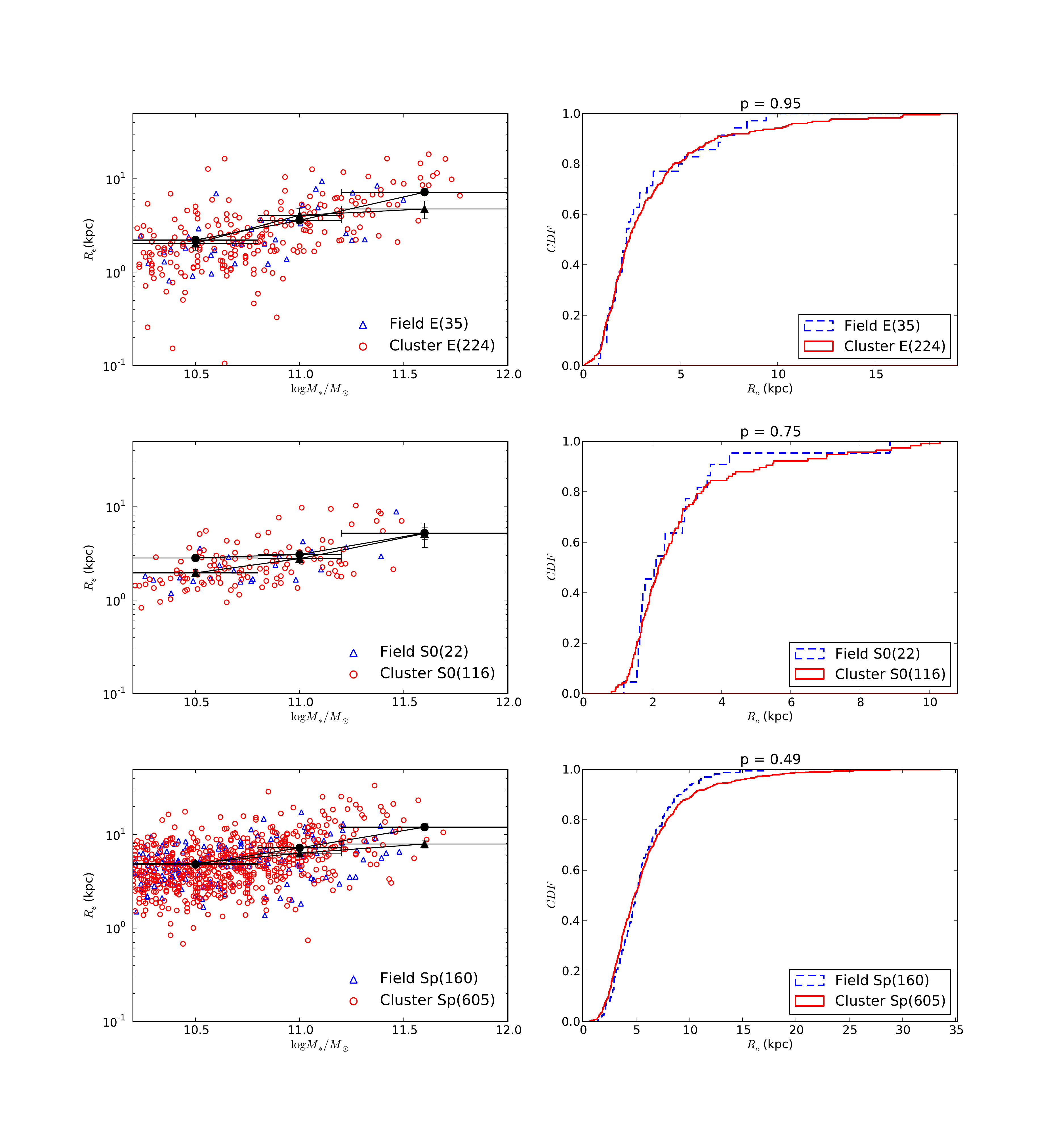}
 \caption{\label{figmsphotmorph} As figure~\ref{figmsspecmorph}, but for the photometric morphology-selected samples.}
\end{figure*}

Figures~\ref{figmsspecmorph} and~\ref{figmsphotmorph} show the stellar mass--size relation for the spectroscopic and photometric samples divided by morphology into Spirals, Ellipticals and S0s. K--S tests comparing the cumulative size distributions of field and cluster galaxies with the same morphology indicate that the probability that the field and cluster samples have the same $R_{\rm e}$ distribution is always over $\sim50$\% ($<1\sigma$ significance). This clearly indicates that we have not detected any significant effect of the galaxy environment on galaxy sizes. This result is in agreement with the work of \citet{maltby10} who found that at $z\sim0.167$ no significant environmental effect was evident on the sizes of cluster and field galaxies of a given morphology. Reassuringly, we obtain entirely consistent results with the spectroscopic sample, smaller but more robust, and with the photometric sample, larger but with smaller reliability for individual galaxies.    

Finding no significant differences in the galaxy size distributions in clusters and the field implies either that such differences do not exist or that, if they do, they are too small to be detected in our sample. In order to estimate how large a difference has to be for us to be able to detect it with our data, we built artificial field samples by randomly increasing the sizes of the cluster galaxies by different average amounts and compared these artificial samples with the original cluster samples via K--S tests. This procedure ensures that the numbers of field and cluster galaxies compared are the same as in the original tests. Moreover, by construction, the simulations use the observed size distributions of the cluster and field samples, thus retaining any putative intrinsic differences. We concluded that we would have been able to detect an average size difference of $\sim20$\% ($\sim10$\%) in $R_{\rm e}$ with $2\sigma$ significance had it been present in the photometric (spectroscopic) sample. We therefore rule out size differences of such magnitude between field and cluster galaxies of a given morphology.

\subsection[]{Mass--size relation for colour-selected samples}
\label{subsecmscol}

\begin{figure*}
 \includegraphics[width=1.1\textwidth]{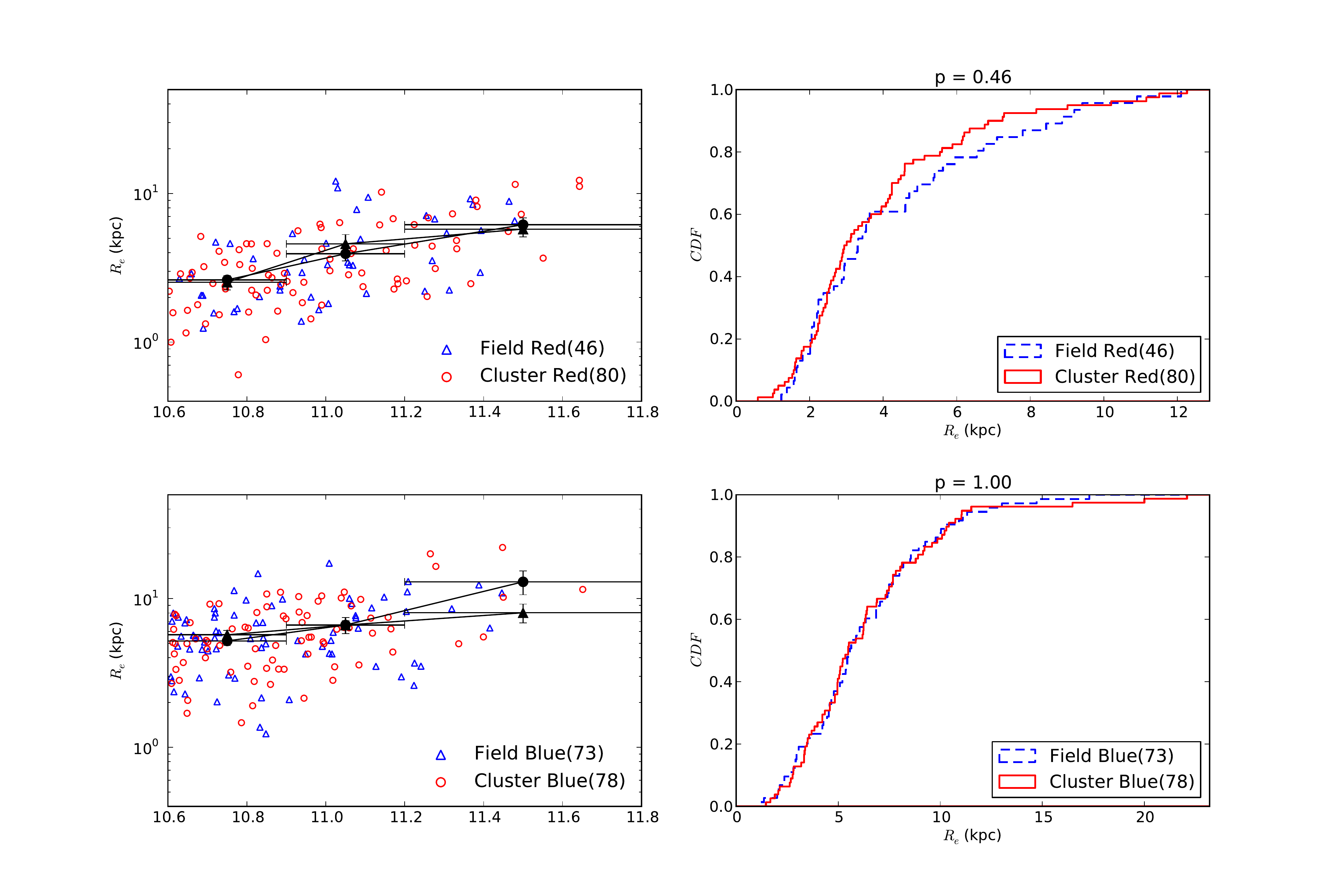}
 \caption{\label{figmsspeccol} Stellar mass--size relation for the spectroscopic colour-selected samples. {\textit{Left:}} The stellar mass--size relations for blue and red field (blue triangles) and cluster (red circles) galaxies. The larger black triangles and circles show average values in arbitrary mass bins for the field and cluster galaxies respectively. The horizontal error bars span the size of each bin, and the vertical error bars correspond to $1\sigma$-errors in the mean $R_{\rm e}$ of each bin.
{\textit{Right:}} The corresponding $R_{\rm e}$ cumulative distribution function (CDF) for the red and blue samples. The dashed blue and red lines correspond to the field and cluster samples respectively.   The numbers above each CDF plot correspond to the $p$ value derived form the K-S tests discussed in the text. The numbers in the boxes denote sample sizes.}

\end{figure*}

\begin{figure*}
 \includegraphics[width=1.1\textwidth]{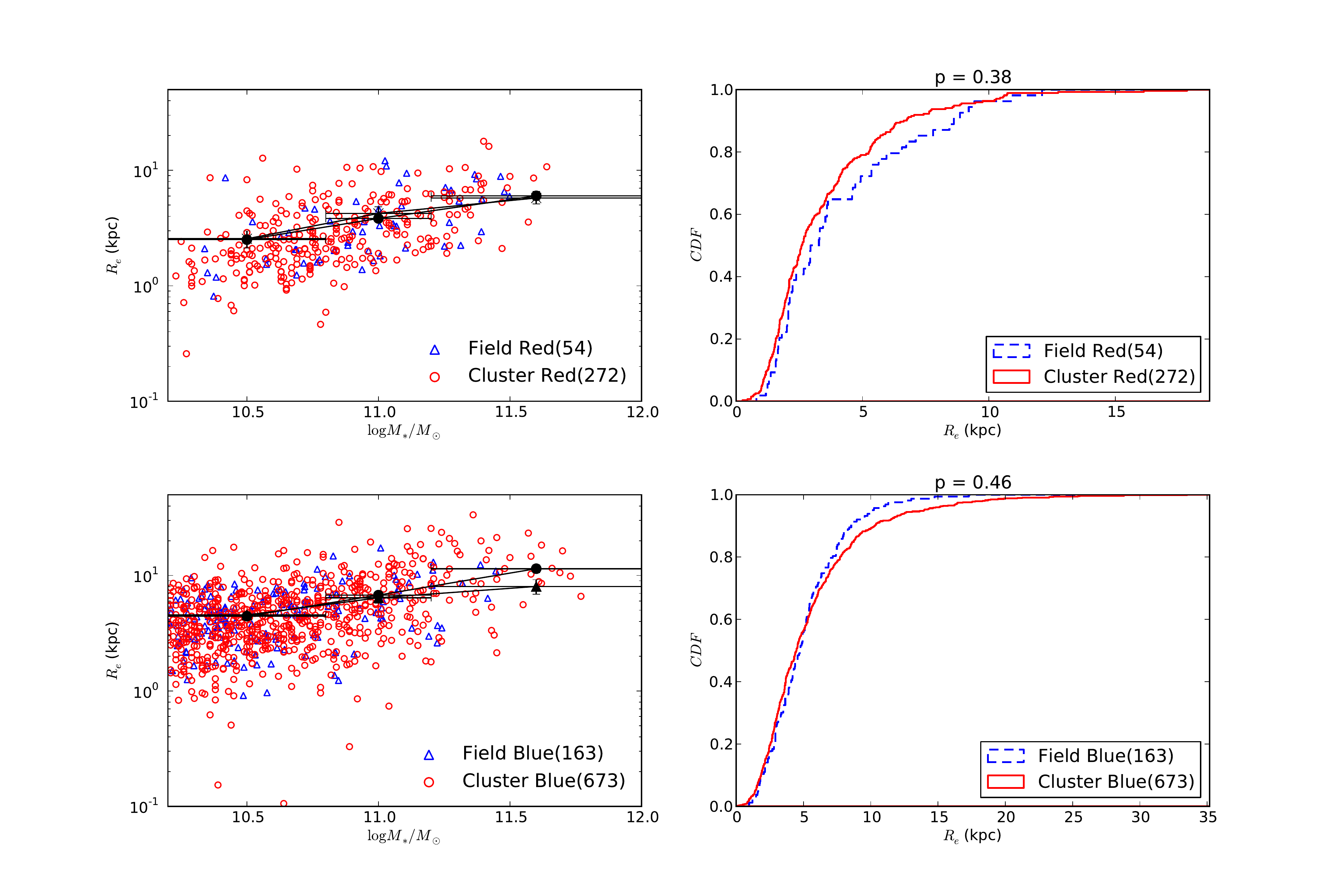}
 \caption{\label{figmsphotcol} As figure~\ref{figmsspeccol}, but for the photometric morphology-selected samples.}
\end{figure*}

Having found no significant difference between the mass--size relations of cluster and field galaxies with similar morphologies, it is interesting to carry out a parallel exercise dividing the galaxy samples by colour since several studies use colour selection when reliable morphological or structural information is not available \citep[e.g., ][]{lani13}. The galaxies were divided into red and blue subsamples taking into consideration their position with respect to the red sequence in a rest-frame $B-V$ colour-magnitude diagram, as described in Section~\ref{subsectcolour}. Figures~\ref{figmsspeccol} and~\ref{figmsphotcol} show the stellar mass--size relation for the spectroscopic and photometric samples of field and cluster galaxies divided by colour, together with their respective cumulative size distributions. As before, K--S tests fail to detected any significant effect of the cluster environment on galaxy sizes. Entirely consistent results are derived for the spectroscopic and photometric samples. These results are also robust against reasonable modifications of the red-blue boundary. We would have been able to detect an average size difference of $\sim20$\% ($\sim10$\%) in $R_{\rm e}$ with $2\sigma$ significance had it been present in the photometric (spectroscopic) sample. We therefore rule out size differences of that magnitude between field and cluster galaxies of a given colour. 

\subsection[]{The effect of the cluster and galaxy masses}
\label{clustermass}

{Given the broad range of velocity dispersions (and thus masses) that the clusters in our sample have (Table~\ref{table1}) we tested whether the cluster mass has any effect on our results. We used the velocity dispersion ($\sigma_{cl}$) of the clusters as a proxy for cluster mass and subdivided our sample into low- and high-mass clusters at the median $\sigma_{cl}$ value. We then compared the mass--size relations for galaxies in high- and low-mass clusters with those of field galaxies, divided by morphology and colour. This exercise was repeated for the spectroscopic and photometric samples. Bearing in mind that the cluster samples are halved in this comparison, thus reducing the sensitivity of the tests, we found no significant cluster--field differences, regardless of the cluster mass.}

{Finally, we tested whether the galaxy mass affects our findings by splitting all the galaxy samples in half at the median stellar mass. Once again, we found no statistically-significant difference in the size distributions of field and cluster galaxies for any of the subsamples. }

\section{Conclusions}
\label{secconclusions}

We have examined the mass--size relations of cluster and field galaxies from the ESO Distant Cluster Survey (EDisCS) in the $0.4<z<0.8$ redshift range. We divided the galaxies into cluster and field environments using both spectroscopic and photometric redshifts. The stellar mass range of the spectroscopic sample is $10.6 \le \log M_\ast/M_{\odot} \le 11.8$, while the photometric sample had stellar masses in the  $10.2 \le \log M_\ast/M_{\odot} \le 12.0$ range. By comparing galaxies in the widest possible range of available environments we have been able to assess whether the environment has any effect on the sizes of galaxies with similar morphologies or colours. Our main conclusions are:

\begin{description}
 \item[$\bullet$] We find no significant difference in the size distributions of cluster and field galaxies of a given morphology (E, S0 and Spiral). We rule out average size differences larger than $10$--$20$\% between field and cluster galaxies of similar morphology and mass.
\item[$\bullet$] Similarly, we find no significant difference in the size distributions of cluster and field galaxies of similar rest-frame $B-V$ colours (red sequence and blue cloud). Once again, we rule out average size differences larger than $10$--$20$\%. 
\end{description}

We obtain entirely consistent conclusions with the spectroscopic sample, smaller but more robust, and with the photometric sample, larger but with smaller reliability for individual galaxies. {These results apply to the full range of masses  explored in this paper. Moreover, our findings do not depend on cluster velocity dispersion (or cluster mass). }

These results have important consequences for the physical process(es) responsible for the size evolution of galaxies, in particular the effect of environment on such evolution. The remarkable growth in size observed from $z\sim2.5$ to the present (by a factor of 2--3 for disks and 4--5 for spheroids) has been reported to depend on the environment at higher redshifts ($z>1$), particularly for early-type/passive/red galaxies. {For instance, \citet{cooper11} find that early-type galaxies in higher density regions tend to have 25\% larger effective radii than their counterparts of equal stellar mass and S\'ersic index in lower density environments. The stelar mass range explored by these authors is comparable with ours.  Similarly, \citet{lani13} find that the most massive ($M_\ast>2\times10^{11}M_{\odot}$) quiescent galaxies at $1 < z < 2$ are on average $\sim50$\% larger in high-density environments. Although we cover a broader stellar mass range than Lani et al., the most massive galaxies in our sample do not show such environmental-dependent size differences. 

It is important to point out that the current evidence for an environmental dependence of the mass--size relation at high-$z$ is often limited by small samples and large uncertainties, resulting sometimes in contradictory results \citep[e.g., ][]{saracco14,newman14}. Regardless, our results indicate that large environmentally driven differences in the mass--size relation are not present below $z\sim0.8$--$1$. Our study complements the work of \citet{hcompany13}, who found very similar results when comparing field and group galaxies at similar redshifts and stellar masses. The main difference between both studies is that our sample contains a much larger fraction of massive clusters, while their group and cluster sample is dominated by lower-mass systems. The combined results of both pieces of research clearly show that at $z<1$ galaxies with a broad range of stellar masses, morphologies and environments show mass--size relations which are independent of environment (at fixed morphologies and/or colours). Interestingly, using a relatively large sample of $\sim400$ early-type galaxies in clusters at $0.8 < z < 1.5$, \citet{delaye14} found that the size distribution of cluster early-types is skewed towards larger sizes compared with that of the field. This results in the average size of cluster early-types being $\sim30$--$40$\% larger, while the median size is similar in clusters and in the field. We do not find such difference in our sample, which spans a lower redshift range. If both results are correct, it would indicate that the transition epoch when the field galaxy size distribution has become similar to the cluster one is somewhere in the region $z\sim0.8$--$1$, the redshift boundary between both works. This is, of course, highly speculative, and it would require a homogeneous study spanning the full redshift range of these two works to confirm this.  
The combination of all these results imply that, if the reported size difference at higher-$z$ is real, the size growth of field galaxies has caught up with that of the cluster galaxies by $z\sim0.8$--$1$. Any putative mechanism proposed as responsible for galaxy growth has to account for the existence of environmental differences at high redshift, if these are confirmed to be real, and their absence (or significant weakening) at lower redshifts. Hierarchical models of galaxy evolution tend to predict moderate-to-strong environmental dependence of galaxy sizes, with the median size increasing by a factor of $\sim1.5$--$3$ when moving from low- to high-mass host haloes \citep[see, e.g.,][]{shankar14}. Since such dependence has been ruled out by observations at almost all redshifts, significant improvement is still needed on the modelling front. Furthermore, larger and more robust observational studies of the galaxy size evolution at all redshifts and in all environments are clearly needed to reduce the uncertainties and tighten the constraints.

\section*{Acknowledgments}

KK would like to thank Boris H\"{a}u\ss{}ler for the useful discussions regarding galaxy fitting and GALAPAGOS. MEG was supported by an STFC Advanced Fellowship. 
KK would also like to thank the referee for the useful comments and suggestions that helped in making the content of this paper better.
BV was supported by the World Premier International Research Center Initiative (WPI), MEXT, Japan and by the Kakenhi Grant-in-Aid for 
Young Scientists (B)(26870140) from the Japan Society for the Promotion of Science (JSPS). 

Based on observations made 
with the NAS/ESA \textit{Hubble Space Telescope}, obtained at the Space Telescope Science Institute, which is operated by the Association of Universities for Research in 
Astronomy, Inc., under NASA contract NAS 5-26555. These observations are associated with proposal 9476. Support for this proposal was provided by NASA through grant
from the Space Telescope Science Institute.

Based on observations obtained at the ESO Very Large Telescope (VLT) as a part of the Large Programme 
166.A-0162

\bibliographystyle{mn2e}
\bibliography{kk_project1}

\appendix

\bsp

\label{lastpage}

\end{document}